\documentclass[aps,prl,floatfix,superscriptaddress,twocolumn,footinbib]{revtex4-2}
\usepackage{graphicx}
\usepackage{dcolumn}
\usepackage{bm}
\usepackage{makecell}
\usepackage{color}

\begin{document}

\title{Catalogue of topological electrons and phonons in all allotropes of carbon}

\author{Qing-Bo Liu}
\affiliation{Hubei Key Laboratory of Optical Information and Pattern Recognition, School of Optical Information and Energy Engineering, School of Mathematics and Physics, Wuhan Institute of Technology, Wuhan, 430073, China}
\author{Xiang-Feng Yang}
\author{Zhe-Qi Wang}
\affiliation{School of Physics and Wuhan National High Magnetic Field Center,
Huazhong University of Science and Technology, Wuhan 430074, People's Republic of China.}
\affiliation{Institute for Quantum Science and Engineering,
 Huazhong University of Science and Technology, Wuhan, Hubei 430074, China.}
\author{Ziyang Yu}
\altaffiliation{tommyu91@163.com}
\author{Lun Xiong}
\altaffiliation{xionglun@wit.edu.cn}
\affiliation{Hubei Key Laboratory of Optical Information and Pattern Recognition, School of Optical Information and Energy Engineering, School of Mathematics and Physics, Wuhan Institute of Technology, Wuhan, 430073, China}
\author{Hua-Hua Fu}
\altaffiliation{hhfu@hust.edu.cn}
\affiliation{School of Physics and Wuhan National High Magnetic Field Center,
Huazhong University of Science and Technology, Wuhan 430074, People's Republic of China.}
\affiliation{Institute for Quantum Science and Engineering,
 Huazhong University of Science and Technology, Wuhan, Hubei 430074, China.}

\date{\today}

\begin{abstract}

Carbon, as one of the most common element in the earth, constructs hundreds of allotropic phases to present rich physical nature. In this work, by combining the ab inito calculations and symmetry analyses method, we systematically study a large number of allotropes of carbon (703), and discovered 315 ideal topological phononic materials and 32 topological electronic materials. The ideal topological phononic nature includes single, charge-two, three, four Weyl phonons, the Dirac or Weyl nodal lines phonons, and nodal surfaces phonons. And the topological electron nature includes topological insulator, (Type-II) Dirac points, triple nodal points, the Dirac (Weyl) nodal lines, quadratic nodal lines and so on. For convenience, we take the $uni$ in SG 178 and $pbg$ in SG 230 as the examples to describe the topological features in the main. We find that it is the coexistence of single pair Weyl phonons and one-nodal surfaces phonons in the $uni$ in SG 178, which can form the single surface arc in the (100) surface BZ and isolated double-helix surface states (IDHSSs)in the (110) surface BZ. In topological semimetal $pbg$ in SG 230, we find that the perfect triple degenerate nodal point can be found in the near Fermi level, and it can form the clear surface states in the (001) and (110) surface BZ. Our work not only greatly expands the topological features in all allotropes of carbon, but also provide many ideal platforms to study the topological electrons and phonons.

\end{abstract}
\maketitle
{\bf{1. Introduction.}}

Topological materials, including topological insulators (TIs), topological semimetals (TSMs) {\color{blue}\cite{1,2,3,4,5,6,7,8,9}}, topological superconductors (TSCs) {\color{blue}\cite{9,10,11,12}} and so on, have been received intense studies due to their protected boundary states and prospects for the future applications in quantum devices in the past a dozen years decades. And nodal point phonons, nodal lines phonons and nodal surface phonons have all been attracting the tremendous research interests, because the topological phonons can realize particular phonon-based devices {\color{blue}\cite{13,14}}, such as phonon diode effect in hexagonal honeycomb lattice. Recently years, single Weyl phonons with Chern number $\pm$1 {\color{blue}\cite{15,16,16_1,16_2,16_3,16_4}}, charge-2 {\color{blue}\cite{17,18}}, 3 {\color{blue}\cite{19,20}} and 4 {\color{blue}\cite{3,21}} Weyl phonons with charges of $\pm$2, $\pm$3 and $\pm$4 have been observed in the realist materials and they can form the single, double, triple and quadruple-helicoid surface states. At the same time, the spin-1 Weyl phonons, charge-2 Dirac phonons and helicoid nodal lines phonons are all observed in the experiment {\color{blue}\cite{22,23}}. So, to explore novel topological electrons and phonons has been another central topic in topological physics and materials.

Besides the well-known the allotropes of carbon of  graphite {\color{blue}\cite{24}}, diamond {\color{blue}\cite{25}}, carbon nanotubes {\color{blue}\cite{26}}, graphene {\color{blue}\cite{27}} and fullerenes {\color{blue}\cite{28}},  more than 700 carbon allotropes have been theoretically predicted or experimentally synthesized. In recent years, the magic angle twisted bilayer graphene (TBG) has attracted the an intense interest of researchers, because it has many interesting physical properties, such as superconductivity  {\color{blue}\cite{29,30,31}} and topologically nontrivial electronic states. Beyond TBG, the topological features of carbon allotropes have been widely studied in the electronic and phononic systems {\color{blue}\cite{32,33,34}}, For example, topological semimetals phases have been predicted in nanostructure carbon allotropes, body centered orthorhombic C$_{16}$ {\color{blue}\cite{35}}  monoclinic C$_{16}$ {\color{blue}\cite{36}} and C$_{40}$ {\color{blue}\cite{37}}, and others {\color{blue}\cite{38,39}}, including the topological features of nodal points {\color{blue}\cite{38}}, nodal nets {\color{blue}\cite{37,38}}, nodal rings {\color{blue}\cite{32}} and nodal surfaces {\color{blue}\cite{39}}. And topological phonons are also predicted in the allotropes of carbon, such as nodal rings phonons {\color{blue}\cite{40}}, straight nodal lines phonons {\color{blue}\cite{41}}, single Weyl phonons {\color{blue}\cite{41}}, charge-2 Weyl phonons {\color{blue}\cite{41}}. However, no reports have reported the all topological electrons and phonons of all allotropes of carbon so far.

\begin{figure*}
\includegraphics[width=16.8 cm]{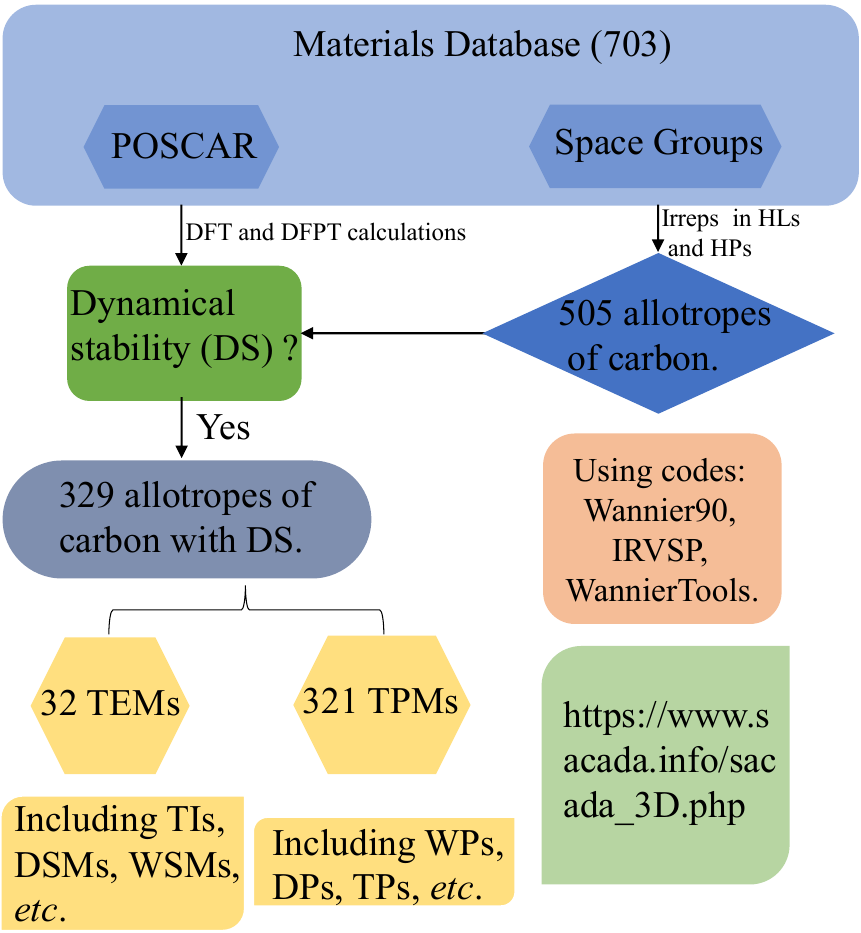}
\caption{The schematic procedure for discovering the topological electrons and phonons of all carbon allotropes.}
\end{figure*}

In this work, by combining the ab initio calculation and symmetry analyses method,  we firstly systematically study a large number of carbon allotropes (703), and discovered 315 ideal topological phononic materials and 32 topological electronic materials. The ideal topological phononic features include charge-one, two, three, four Weyl phonons, Dirac or Weyl phonons, Dirac or Weyl nodal lines phonons, and nodal surfaces phonons. And the topological electron nature includes topological insulators, Dirac or Weyl points, triple nodal points, the Dirac or Weyl nodal lines, quadratic nodal lines in all allotropes of carbon.  We mainly discuss the topological phononic (electronic) nature of $uni$ ($pbg$) in SG 178 (230) in the main. We find that it is the coexistence of single pair Weyl phonons with charge -1 and one-nodal surfaces phonons in the $uni$ in SG 178. Moreover, it can form the single surface arc in the (100) surface BZ and isolated double-helix surface states in the (110) surface BZ. In addition, another example of topological semimetal $pbg$ in SG 230 can form the ideal triple nodal point near the Fermi level. And it can also  form clear surface states in (001) and (110) surfaces states. We discuss the topological electronic and phononic features of other carbon allotropes in the  Supplementary materials {\color{blue}\cite{41_1}}. Our theoretical results not only find many ideal platform to study the topological electron and phonons in allotropes of carbon, but also greatly expands the topological features in all carbon allotropes.

\begin{figure*}
\includegraphics[width=17.5 cm]{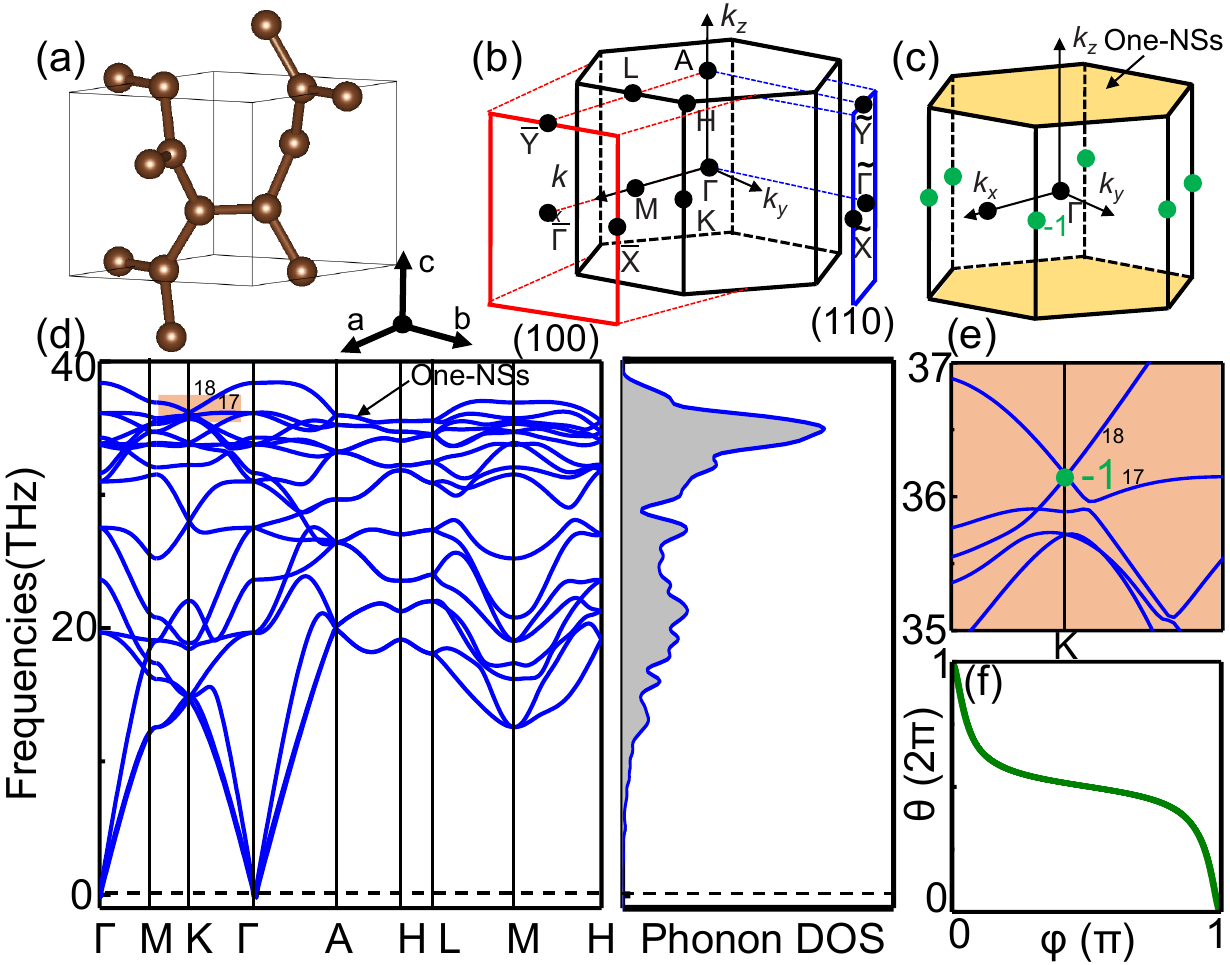}
\caption{(a) Crystal structure of $uni$ in SG 178 in a primitive cell. (b) The bulk BZ of $uni$ and the (100) (red square) and (110) (blue square) surface BZ. (c) The distribution of single pair monopole WPs (green dots with Chern number -1) and one-nodal surfaces (yellow region) between the $17^{th}$ and $18^{th}$ phonon bands of $uni$ in the first BZ. (d) The phonon spectra of $uni$ along the high-symmetry paths and the corresponding phononic density of states (DOS). A red box is marked along the high-symmetry paths as shown in Fig. 1(e). (f) The evolution of the average position of Wannier centers for single pair monopole WPs.}
\end{figure*}

{\bf{2. Materials and method}}

The crystallographic data of all allotropes of carbon can be taken from the international database SACADA {\color{blue}\cite{42}}. The phononic dispersions and elecronic bands of allotropes of carbon are calculated by the density functional theory (DFT) using the Vienna \emph{ab initio} Simulation Package (VASP) with the generalized gradient approximation (GGA) in the form of Perdew-Burke-Ernzerhof (PBE) function for the exchange-correlation potential {\color{blue}\cite{43,44,45}}. An accurate optimization of structural parameters is employed by minimizing the interionic forces less than 0.001 $\textrm{eV}/\textrm{{\AA}}$ and an cutoff energy at 450 eV. The BZ is gridded with 3$\times$3$\times$3 \emph{k} points. Then the phononic spectra are gained using the density functional perturbation theory (DFPT), implemented in the Phonopy Package {\color{blue}\cite{46}}. The force constants are calculated using a $2\times2\times2$ supercell. To reveal the topological nature of phonons, we construct the phononic Hamiltonian of tight-binding (TB) model and the surface local density of states (DOS) with the open-source software Wanniertools code {\color{blue}\cite{47}} and the surface Green's functions {\color{blue}\cite{48}}. The Chern numbers or topological charge of WPs are calculated by Wilson loop method {\color{blue}\cite{49}}. The electronic surface states have been performed using the open-source code WANNIERTOOLS based on the Wannier tight-binding model constructed using the WANNIER90 code. The irreps of the electronic and phononic bands are computed by the program {\color{blue}$ir2tb$} on the electronic and phononic Hamiltonian of TB model {\color{blue}\cite{50}}.

{\bf{3. Results and discussion}}

{\emph{3.1 Research the topological electrons and phonons}}

The guiding principle of our search the topological electrons and phonons in all carbon allotropes as shown in Fig. 1. Firstly, among 703 carbon allotropes structures in the international database SACADA {\color{blue}\cite{42}}, we will exclude the no topological features of space groups, according to the irreducible representations of high-symmetry lines (HLs) and high-symmetry points (HPs) in space groups {\color{blue}\cite{51,52}}. Then, we can obtain the 505 allotropes of carbon which can exist the topologically electronic or phononic properties. Secondly, their dynamical stabilities were examined by using the density functional perturbation theory (DFPT) calculations, which yields 329 dynamical stable allotropes of carbon. Finally, by using the first-principle calculations, we can divide 329 dynamical stable allotropes of carbon into 32  perfect topological electronic materials (TEMs) including topological insulators (TIs), Dirac semimatals (DSMs), Weyl semimatals (WSMs) and so on, and 321  perfect topological phononic materials (TPMs) including Weyl phonons (WPs), Dirac phonons (DPs), triple-fold phonons (TPs) and so on as shown in Supplementary materials {\color{blue}\cite{41_1}}.

{\emph{3.2 Symmetry analysis}}
\begin{table}[!t]
\caption{The complete list of allotropes of carbon with topological properties in SGs 178 and 230. The first column indicates the space groups (SGs), the second column indicates the number of allotropes of carbon in SACADA, the third column indicates the SG symbol, and the fourth column shows the types of topological features (TFs). The red letters stand for the topological insulators or semimatel.
}\label{tab:CFWPs}
\begin{tabular}{p{2cm}p{2cm}p{2cm}p{2.2cm}}
\hline
\hline
\makecell[c]{SGs}   & \makecell[c]{Numbers}    & \makecell[c]{Names}  &\makecell[c]{TFs}  \\
\hline
\makecell[c]{P6$_1$22(No. 178)}  & \makecell[c]{29 \\ 56}   & \makecell[c]{uni \\ unj} &   \makecell[c]{ Weyl points and \\ nodal  surfaces} \\
\hline

\makecell[c]{Ia$\bar{\text{3}}$d(No. 230)}  & \makecell[c]{13 \\ {\color{red}53}}  & \makecell[c]{lcs \\ {\color{red}pbg}}  &  \makecell[c]{triple nodal \\ points}\\
\hline
\hline
\end{tabular}
\end{table}

To elucidate the topologically nontrivial features of single pair Weyl phonons and one-nodal surface phonons in SG 178 in Table-I, we can prove that it exist the single pair Weyl phonons at high-symmetry point K  and one-nodal surface phonons at $k_z$= $\pm\pi$ planes by using a two-band $k\cdot{p}$ model as
\begin{equation}
\mathcal{H}_{kp}(k)=g_x(k)\sigma_x+g_y(k)\sigma_y+g_z(k)\sigma_z,
\end{equation}
where $k=(k_x,k_y,k_z)$, $\sigma_{x,y,z}$ represents the three Pauli matrices, and $g_{x,y,z}(k)$ represents the complex functions versus $k_x$, $k_y$ and $k_z$.

Let us first consider a three screw symmetry $\{C_{31}^+|00\frac{1}{3}\}$ and a two-fold screw symmetry $\{C_{21}^{''}|00\frac{1}{2}\}$ at K point in SG 178. Under a 2D irreps:R3 {\color{blue}\cite{51,52}}, the representation matrixes of $\{C_{31}^+|00\frac{1}{3}\}$ and $\{C_{21}^{''}|00\frac{1}{2}\}$ can be described as

\begin{eqnarray*}
\{C_{31}^+|00\frac{1}{3}\}=\left[\begin{array}{ll}
-\frac{1}{2} & \frac{\sqrt{3}}{2} \\
-\frac{\sqrt{3}}{2} & -\frac{1}{2}
\end{array}\right], \\
\{C_{21}^{''}|00\frac{1}{2}\}=\left[\begin{array}{ll}
0 & 1 \\
1 & 0
\end{array}\right],\\
\end{eqnarray*}

So,the $k\cdot{p}$-invariant Hamiltonian at point K is derived as
\begin{eqnarray*}
\mathcal{H}_{kp}=ak_{x}\sigma_{x}+bk_{z}\sigma_{y}+ck_{y}\sigma_{z},
\end{eqnarray*}
where a, b and c are constant coefficients. According to the two-band $k\cdot{p}$  Hamiltonian, we find that it has a $k$ dispersion along all phononic dispersion's direction at K point.

Then, we will prove that it can exist the one nodal surfaces at at $k_z$= $\pm\pi$ planes in the nonsymmorphnic SG 178. We continue to consider the skew axial symmetry $S_{2z}$: $(x,y,z,t) \mapsto (-x,-y,-z+\frac{1}{2},t)$, which indicates that $S_{2z}$ inverses $k_x$ and $k_y$ while preserves $k_z$ in the \emph{k}-space. One can drive that $(S_{2z})^2$ = $T_{001}$ = $e^{-ik_{z}}$, where $T_{001}$ is the translation along the $x$ direction by a lattice constant. As we know that $\mathcal{T}$ is antiunitary and inverses $k$ with the relation $T^2$ = 1. So the compound symmetry $S$ = $T$$S_{2z}$ is antiunitary and only inverses $k_z$. Since [$T$, $S_{2z}$] =0, $S$ satisfies
\begin{equation}
S^2=e^{-ik_{z}}.
\end{equation}

The crossing on the surface can be comprehended as a result of the Kramer's degeneracy. Under the operation $S$, we find that any point on the $k_z = \pi$ is invariant. Moreover, from Eq. (2), one can drive that the antiunitary symmetry satisfies $S^2$ = -1 on the whole $k_z = \pi$ plane. Thus, the two Kramer's degeneracy can arise at the every point in this plane. Away from this plane, this Kramer's degeneracy is generically destroyed owing to the loss of symmetry protections. In a word, a nodal surface should be formed at the $k_z = \pi$ plane.

\begin{figure*}
\includegraphics[width=17.5 cm]{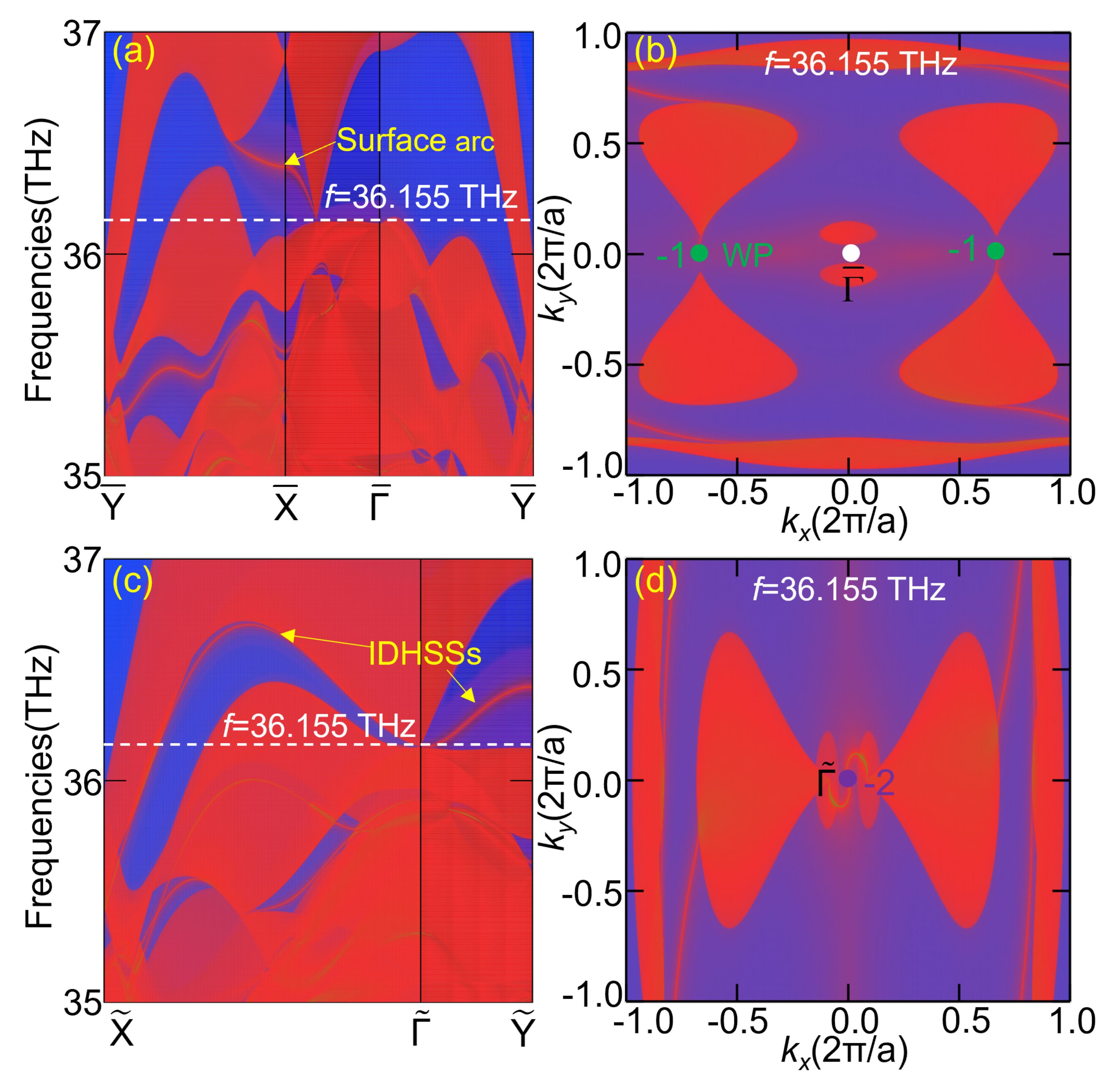}
\caption{The surface phonon dispersions and isofrequency surface contours of $uni$ in SG 178. (a) The surface phonon dispersion on the (100) surface along the high-symmetry $\overline{\textrm{Y}}$-$\overline{\textrm{X}}$-$\overline{\textrm{$\Gamma$}}$-$\overline{\textrm{Y}}$. (b) The isofrequency surfaces contours of (100) surface BZ at 36.155 THz. (c) The surface phonon dispersion on the (110) surface along the high-symmetry $\widetilde{\textrm{X}}$-$\widetilde{\textrm{$\Gamma$}}$-$\widetilde{\textrm{Y}}$. (d) The isofrequency surfaces contours 0f (110) surface BZ at 36.155 THz.}

\end{figure*}

Next, we will derive the $k\cdot{p}$ model of $\Gamma$ point in SG 230 in Table-I. The point $\Gamma$ belongs to point group $O_h$, which includes five symmetries $\{S_{61}^{-}|000\}$, $\{\sigma_{x}|\frac{1}{2}\frac{1}{2}0\}$, $\{\sigma_{z}|\frac{1}{2}0\frac{1}{2}\}$, $\{C_{2c}|00\frac{1}{2}\}$ and time-reversal symmetry $\mathcal{T}$. Under a 3D irreps:$\Gamma_4^+$, the representation matrixes of there symmetries  can be shown

\begin{eqnarray*}
\{S_{61}^{-}|000\}=\left[\begin{array}{lll}
0 & 0 & 1 \\
1 & 0 & 0 \\
0 & 1 & 0 \\
\end{array}\right], \\
\{\sigma_{x}|\frac{1}{2}\frac{1}{2}0\}=\left[\begin{array}{lll}
-1 & 0 & 0 \\
0 & 1 & 0 \\
0 & 0 & -1
\end{array}\right],\\
\{\sigma_{z}|\frac{1}{2}0\frac{1}{2}\}=\left[\begin{array}{lll}
0 & 1 & 0 \\
1 & 0 & 0 \\
0 & 0 & -1
\end{array}\right],\\
\{C_{2c}|00\frac{1}{2}\}=\left[\begin{array}{lll}
-1 & 0 & 0 \\
0 & 1 & 0 \\
0 & 0 & -1
\end{array}\right],\\
\mathcal{T} = \mathcal{K}.
\end{eqnarray*}
The $\mathcal{K}$ is a complex conjugate operator. So, the three-band $k\cdot{p}$-invariant Hamiltonian at point $\Gamma$ is derived as
\begin{eqnarray*}
\mathcal{H}_{kp}=\left[\begin{array}{lll}
0 & mk_{x}k_{y} & nk_{x}k_{z} \\
mk_{x}k_{y} & 0 & qk_{y}k_{z} \\
nk_{x}k_{z} & qk_{y}k_{z} & 0
\end{array}\right],\\
\end{eqnarray*}
where m, n and q are real constant coefficients. According to this equation, we find that it can form quadratic triple nodal point (QDNP) at the $\Gamma$ point in SG 230.

{\emph{3.3 Topological phononic features of $uni$ in SG 178 }}

The crystal structure of $uni$ in SG 178 is shown in Fig. 2(a), which includes 6 carbon atoms in a unit cell. The bulk BZ, (110) surface BZ (blue square) and (100) surface BZ (red square) are shown in Fig. 2(b). The phononic bands and the corresponding phononic density of states (DOS) of $uni$ are drawn in Fig. 2(d), where no imaginary frequencies indicate that this material is thermodynamically stable. From the first glance, we can find that two single pair Weyl points are located at the K point formed by 17$^{th}$ and 18$^{th}$ bands, and their distribution of single pair Weyl phonons (green dots) are shown in Figs. 2(c) and 2(e) with Chern number -1 in Fig. 2(f). And two-fold degeneracy bands can be observed along the high-symmetry lines A-H-L in Fig. 2(d), which can form one nodal surfaces at $k_z =\pm \pi$ planes (yellow region) in Fig. 2(c).

In order to study the topological features of $uni$, we calculate the (100) and (110) surface states of $uni$ in Fig 3. As shown in Fig. 1(b), the high-symmetry points $\Gamma$ and M (L and A) are projected to $\overline{\textrm{$\Gamma$}}$ ($\overline{\textrm{Y}}$) in the (001) surface in Fig. 2(b). In (100) surface, one can see that the surface arcs surface states can be formed by the single Weyl phonons and one nodal surfaces in Fig. 3(a), which is for the first time reported in the phononic systems. We also calculate the sofrequency surface at frequency 36.155 THz and we observe that the isolated surface acrs can be formed by a Weyl phonon with charge of -1 in Fig. 3(b).  In (110) surface BZ, the high-symmetry points $\Gamma$ and K (A) are projected to $\widetilde{\textrm{$\Gamma$}}$ ($\widetilde{\textrm{Y}}$) in Fig. 2(b). We find that two Weyl phonons with charge of -1 are projected o the same point which can form a charge-two Weyl phonons at $\widetilde{\textrm{$\Gamma$}}$, leading to isolated double-helix surface states in Figs. 3(c). And we also can observe the isolated double-helix surface arc states in the sofrequency surface at frequency 36.155 THz  as shown in Fig. 3(d). So, these novel physical features confirm that it exit the topological nontrivial nature in $uni$.

{\emph{3.4 Topological electronic features of $pbg$ in SG 230 }}

Next, we will disscus the topological electronic features of $pbg$ in SG 230 in Figs. 4 and 5. The unit cell of $pbg$ is shown in Fig. 4(a), which includes 48 carbon atoms. The 3D BZ, (001) (blue square) and (110) (red square) surface BZ are shown in Fig. 4(b). The electronic bands and density of states (DOS) of $pbg$ are draw in Fig. 4(c), where blue lines (dashed line) stand for the result of DFT (Wannier90 fitting). We find that a ideal triple nodal point can exist near the Fermi level ($E_f$) at the point $\Gamma$.

Like to $uni$ in SG 178, we also calculate the (001) and (110) surface states of $pbg$ in Fig. 5.  The $\Gamma$ and H points (P and H) are projected to $\overline{\textrm{$\Gamma$}}$ ($\overline{\textrm{M}}$) in the (001) surface in Fig. 4(b).  At the first glance, we find that at the point $\overline{\textrm{$\Gamma$}}$, there are three surface arcs surfaces states projected by triple nodal points in Fig. 5(a). And we can also observe the triple nodal point at the Fermi level as shown in Fig. 5(b). In (110) surface, we can also clearly observe the three surface arcs surfaces states projected by triple nodal points at point $\widetilde{\textrm{$\Gamma$}}$ in Fig. 5(c) and triple nodal point at $\widetilde{\textrm{$\Gamma$}}$ in Fig. 5(d). These nontrivial topological feature all prove that $pbg$ is topological.

\begin{figure*}
\includegraphics[width=17.5 cm]{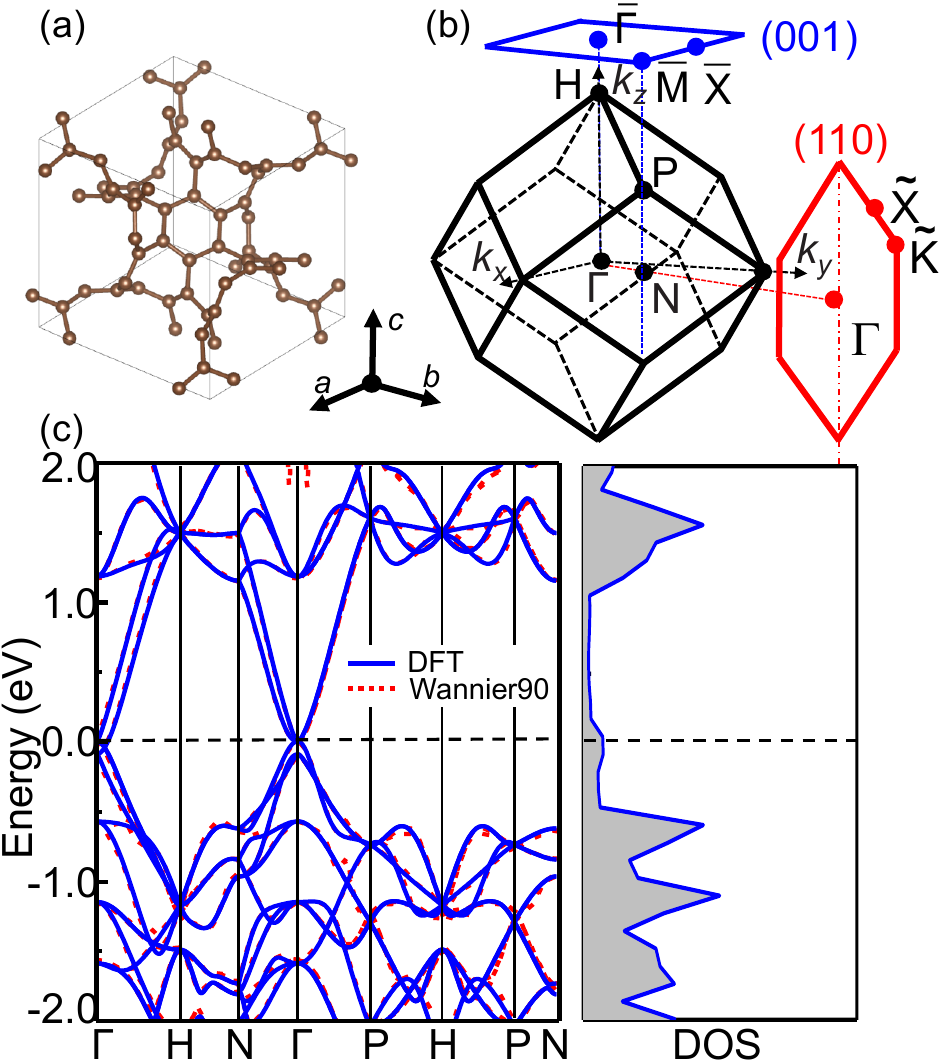}
\caption{(a) Crystal structure of $pbg$ in SG 230 in a primitive cell. (b) The bulk BZ of $pbg$ and the (001) (blue square) and (110) (red square) surface BZ. (c) The DFT (blue solid lines) and Wannier90 (red dashed lines) electronic bands of $pbg$ along the high-symmetry paths and the corresponding electronic density of states (DOS).}
\end{figure*}

\begin{figure*}
\includegraphics[width=17.5 cm]{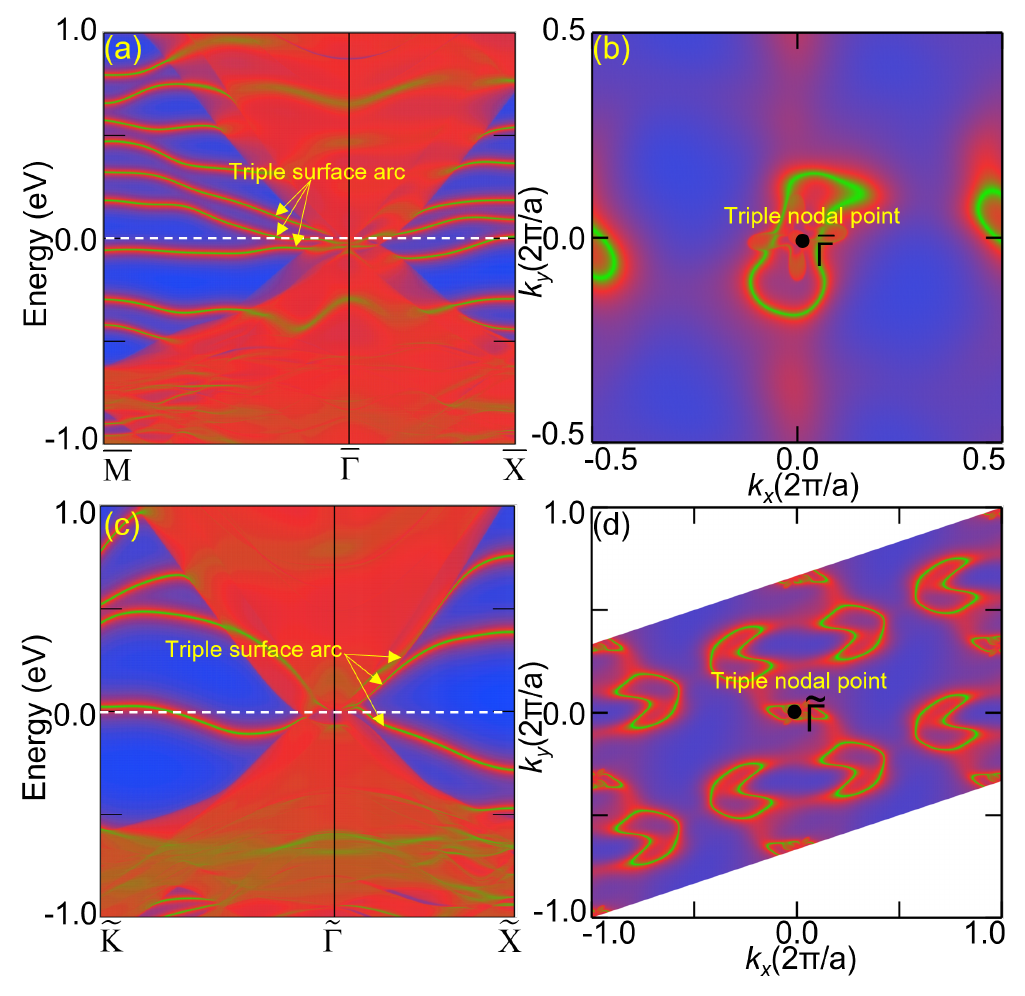}
\caption{The surface electronic dispersions and isofrequency surface contours of $pbg$ in SG 230. (a) The surface phonon dispersion on the (100) surface along the high-symmetry $\overline{\textrm{M}}$-$\overline{\textrm{$\Gamma$}}$-$\overline{\textrm{X}}$. (b) The isofrequency surfaces contours of (001) surface BZ at Fermi level. (c) The surface electronic dispersion on the (110) surface along the high-symmetry $\widetilde{\textrm{K}}$-$\widetilde{\textrm{$\Gamma$}}$-$\widetilde{\textrm{X}}$. (d) The isofrequency surfaces contours 0f (110) surface BZ at Fermi level.}
\end{figure*}\textbf{}

{\bf{4. Conclusion}}

In summary, performing symmetry analysis and first principles calculations, we systematically study 703 allotropes of carbon, and discovered 315 ideal topological phononic materials, including single, charge-two, three, four Weyl phonons, the Dirac or Weyl nodal lines phonons, and nodal surfaces phonons, and 32 topological electronic materials, including topological insulator, (Type-II) Dirac points, triple nodal points, the Dirac (Weyl) nodal lines, quadratic nodal lines and so on. In $uni$ carbon in SG 178, we find that it is the coexistence of single pair Weyl phonons and one-nodal surfaces phonons in the uni in SG 178, which can form the single surface arc in the (100) surface BZ and isolated double-helix surface states in the (110) surface BZ. Another example, we find we find that the perfect triple nodal point can be found in the near Fermi level, and it can form the clear triple surface arc states in the (001) and (110) surface BZ. And more carbon allotropes are tabulated in the Supplementary materials {\color{blue}\cite{41_1}}. Our work not only greatly expands the topological features in all allotropes of carbon, but also provide many ideal platforms to study the topological electrons and phonons.

{\bf{Conflict of interest} }

The authors declare that they have no conflict of interest.

{\bf{Acknowledgements.}}

This work is supported by the National Science Foundation of China with Grants Nos. 11774107, 12147113, 12104348 and U20A2077, by the Science and Technology Department of Hubei Provincial with Grant No. 2022CFD041, and partially by the National Key R\&D Program of China (2021YFC2202300).

{\bf{Author contributions}}

Ziyang Yu, Lun Xiong and Hua-Hua Fu proposed and supervised the project. Qing-Bo Liu carried out the high-throughput calculations. All authors contributed to writing of the manuscript.

{\bf{Author contributions}}

Supplementary materials to this article can be found online at XXXX.


\begin{thebibliography}{apssamp}


\bibitem{1} B. Bradlyn, J. Cano, Z. Wang, M. G. Vergniory, C. Felser, R. J. Cava, and B. A. Bernevig, Begyond Dirac and Weyl ferimions: unconvertianl quasiparticles in conventional crystals, {\color{blue}Science \textbf{353}, 6299 (2016)}.

\bibitem{2} X. Huang, L. Zhang, Y. Long, P. Wang, D. Chen, Z. Yang, H. Liang, M. Xue, H. Weng, Z. Fang, X. Dai, and G. Chen, Obversation of the chiral-anomaly-induced negative magnetoresistance in 3D Weyl semimetal TaAs, {\color{blue}Phys. Rev. X \textbf{5}, 031023 (2015)}.

\bibitem{3} T. T. Zhang, R. Takahashi, C. Fang, and S. Murakami, Twofold quadruple Weyl nodes in chiral cubic crystals, {\color{blue}Phys. Rev. B \textbf{102}, 125148 (2020)}.

\bibitem{4} S. Y. Xu, I. Belopolski, N. Alidoust, M. Neupane, G. Bian, C. L. Zhang, R. Sankar, G. Q. Chang, Z. J. Yuan, C. C. Lee, S. M. Huang, H. Zheng, J. Ma, D. S. Sanchez, B. K. Wang, A. Bansil, F. C. Chou, P. P. Shibayev, H. Lin, S. Jia, and M. Z. Hasan, Discovery of a Weyl fermion semimental and topological Fermi arcs, {\color{blue}Science \textbf{349}, 613 (2015)}.

\bibitem{5} B. Q. Lv, N. Xu, H. M. Weng, J. Z. Ma, P. Richard, X. C. Huang, L. X. Zhao, G. F. Chen, C. E. Matt, F. Bisti, V. N. Strocov, J. Mesot, Z. Fang, X. Dai, T. Qian, M. Shi, and H. Ding, Observation of Weyl nodes in TaAs, {\color{blue}Nat. Phys. \textbf{11}, 724 (2015)}.

\bibitem{6} N. P. Armitge, E. J. Mele, and A. Vishwanath, Weyl and Dirac semimetals in three-dimensional solids, {\color{blue}Rev. Mod. Phys. \textbf{90}, 015001 (2018)}.

\bibitem{7} A. P. Schnyder, S. Ryu, A. Furusaki, and A. W. W. Ludwig, Classification of topological insulators and superconductors in three spatial dimensions, {\color{blue}Phys. Rev. B \textbf{78}, 195125 (2008)}.

\bibitem{8} C. L. Kane and E. J. Mele, Z$_2$ Topological order and the quantum spin Hall effect, {\color{blue}Phys. Rev. Lett. \textbf{95}, 146802 (2005)}.

\bibitem{9} M. Z. Hasan and C. L. Kane, Colloquium: Topological insulators, {\color{blue}Rev. Mod. Phys. \textbf{82}, 3045 (2010)}.

\bibitem{10} R. M. Lutchyn, J. D. Sau, and S. Das Sarma, Majorana Fermions and a topological phase transition in semiconductor-superconductor heterostructures, {\color{blue}Phys. Rev. Lett. \textbf{105}, 077001 2010)}.

\bibitem{11} S. Nadj-Perge, I. K. Drozdov, J. Li, H. Chen, S. Jeon, J. Seo, A. H. MacDonald, B. A. Bernevig, and A. Yazdani, Observation of Majorana fermions in ferromagnetic atomic chains on a superconductor, {\color{blue}Science \textbf{346}, 602}.

\bibitem{12} X. L. Qi and S. C. Zhang, Topological insulators and superconductors, {\color{blue}Rev. Mod. Phys. \textbf{83}, 1057 (2011)}.

\bibitem{13} L. F. Zhang, J. Ren, J. S. Wang, and B. W. Li, Topological nature of the phonon Hall effect, {\color{blue}Phys. Rev. Lett. \textbf{105}, 225901 (2010)}.

\bibitem{14} Y. Liu, X. Chen, and Y. Xu, Topological phononics: from fundamental models to real materials. {\color{blue}Adv. Funct. Mater., \textbf{30}, 1904784 2020}.

\bibitem{15} L. Lu, L. Fu, J. D. Joannopoulos, and M. Soljacic, Weyl points and line nodes in gyroid photonic crystals, {\color{blue}Nat. Photon. \textbf{7}, 294 (2013)}.

\bibitem{16} L. Lu, Z. Y. Wang, D. X. Ye, L. X. Ran, L. Fu, J. D. Joannopoulos, and M. Soljaci, Experimental observation of Weyl points, {\color{blue}Science \textbf{349}, 622 (2015)}.

\bibitem{16_1} Q. Xie, J. Li, S. Ullah, R. Li, L. Wang, D. Li, Y. Li, S. Yunoki, and X.-Q. Chen, Phononic Weyl points and one-way topologically protected nontrivial phononic surface arc states in noncentrosymmetric WC-type materials, {\color{blue}Phys. Rev. B. \textbf{99}, 174306 (2019)}.

\bibitem{16_2} B. W. Xia, R. Wang, Z. J. Chen, Y. J. Zhao, and H. Xu, Symmetry-protected ideal type-II Weyl phonons in CdTe. {\color{blue}Phys. Rev. Lett. \textbf{123}, 065501 (2019)}.

\bibitem{16_3}  G. Ding, F. Zhou, Z. Zhang,  Z. M. Yu, and  X. Wang, Charge-two Weyl phonons with type-III dispersion.  {\color{blue}Phys. Rev. B. \textbf{105}, 134303  (2022)}.


\bibitem{16_4} B. Zheng, B. Xia, R. Wang, Z. Chen, J. Zhao, Y. Zhao, and H. Xu, Ideal type-III nodal-ring phonons, {\color{blue}Phys. Rev. B \textbf{101}, 100303(R) (2020)}.

\bibitem{17} Q.-B. Liu, Y. Qian, H.-H. Fu, and Z. Wang, Symmetry-enforced Weyl phonons, {\color{blue}npj Comput. Mater. \textbf{6}, 95 (2020)}.

\bibitem{18} R. Wang, B.-W. Xia, J. Chen, B.-B. Zheng, J. Zhao, and H. Xu, Symmetry-Protected Topological Triangular Weyl Complex. {\color{blue}Phys. Rev. Lett.  \textbf{124}, 105303 (2020)}.

\bibitem{19} X. Wang, F. Zhou, Z. Zhang, Z.-M. Yu, and Y. Yao, Hourglass charge-three Weyl phonons, {\color{blue}Phys. Rev. B \textbf{106}, 214309 (2022)}.

\bibitem{20} G. Liu, Z. Chen, P. Wu, and H. Xu, Triple hourglass Weyl phonons, {\color{blue}Phys. Rev. B \textbf{106}, 214308 (2022)}.

\bibitem{21} Q.-B. Liu, Z. Wang, and H.-H. Fu, Charge-four Weyl phonons, {\color{blue}Phys. Rev. B \textbf{103}, L161303 (2021)}.

\bibitem{22} T.-T. Zhang, Z. D. Song, A. Alexandradinata, H.-M. Weng, C. Fang, L. Lu, and Z. Fang, Double-weyl phonons in transition-metal monosilicides, {\color{blue}Phys. Rev. Lett. \textbf{120}, 016401 (2018)}.

\bibitem{23} H. Miao, T.-T. Zhang, L. Wang, D. Meyers, A.-H. Said, Y.-L. Wang, Y.-G. Shi, H.-M. Weng, Z. Fang, and M.-P.-M. Dean, Observation of double Weyl phonons in parity-breaking FeSi, {\color{blue}Phys. Rev. Lett. \textbf{121}, 035302 (2018)}.

\bibitem{24} W. Wang, S. Dai, X. Li, J. Yang, D. J. Srolovitz, and Q. Zheng, Measurement of the cleavage energy of graphite, {\color{blue}Nat. Commun. \textbf{6}, 7853 (2015)}.

\bibitem{25} E. A. Ekimov, V. A. Sidorov, E. D. Bauer, N. N. Mel'nik, N. J. Curro, J. D. Thompson, and S. M. Stishov, Superconductivity in diamond, {\color{blue}Nature \textbf{428}, 542-545(2004)}.

\bibitem{26} S. Iijima, Helical microtubules of graphitic carbon, {\color{blue}Nature (London) \textbf{354}, 56 (1991)}.

\bibitem{27} A. H. Castro Neto, F. Guinea, N. M. R. Peres, K. S. Novoselov, and A. K. Geim, The electronic properties of graphene, {\color{blue}Rev. Mod. Phys. \textbf{81}, 109 (2009)}.

\bibitem{28} R. C. Haddon, Chemistry of the Fullerenes: The Manifestation of Strain in a Class of Continuous Aromatic Molecules, {\color{blue}Science  \textbf{261}, 5128 (1993)}.

\bibitem{29} Y. Cao, V. Fatemi, S. Fang, K. Watanabe, T. Taniguchi, E. Kaxiras, and P. Jarillo-Herrero, Unconventional superconductivity in magic-angle graphene superlattices, {\color{blue}Nature \textbf{556}, 43-50(2018)}.

\bibitem{30} Y. Cao, D. Rodan-Legrain, J. M. Park, N. FQ Yuan, K. Watanabe, T. Taniguchi, R. M Fernandes, L. Fu, and P. Jarillo-Herrero, Nematicity and competing orders in superconducting magic-angle graphene, {\color{blue}Science \textbf{372}, 264-271(2021)}.

\bibitem{31} Y. Cao, J. M. Park, K. Watanabe, T. Taniguchi, and P. Jarillo-Herrero, Pauli-limit violation and re-entrant superconductivity in moiré graphene, {\color{blue}Nature \textbf{595}, 526-531(2021)}.

\bibitem{32} H. Weng, Y. Liang, Q. Xu, R. Yu, Z. Fang, X. Dai, and Y. Kawazoe, Topological node-line semimetal in three-dimensional graphene networks, {\color{blue}Phys. Rev. B \textbf{92}, 045108 (2015)}.

\bibitem{33} Y. Chen, Y. Xie, S. A. Yang, H. Pan, F. Zhang, M. L. Cohen, and S. Zhang, Nanostructured Carbon Allotropes with Weyl-like Loops and Points, {\color{blue}Nano Lett. \textbf{15}, 6974 (2015)}.

\bibitem{34} J.-T. Wang, C. Chen, and Y. Kawazoe,Topological nodal line semimetal in an orthorhombic graphene network structure, {\color{blue}Phys. Rev. B \textbf{97}, 245147 (2018)}.

\bibitem{35} J.-T. Wang, H. Weng, S. Nie, Z. Fang, Y. Kawazoe, and C. Chen, Body-Centered Orthorhombic C$_{16}$: A Novel Topological Node-Line Semimetal, {\color{blue}Phys. Rev. Lett. \textbf{116}, 195501 (2016)}.

\bibitem{36} Y. Cheng, X. Feng, X. Cao, B. Wen, Q. Wang, Y. Kawazoe, and P. Jena, Body-Centered Tetragonal C$_{16}$: A Novel Topological Node-Line Semimetallic Carbon Composed of Tetrarings, {\color{blue}Small \textbf{13}, 1602894 (2017)}.

\bibitem{37} J.-T. Wang, S. Nie, H. Weng, Y. Kawazoe, and C. Chen, Topological Nodal-Net Semimetal in a Graphene Network Structure, {\color{blue}Phys. Rev. Lett. \textbf{120}, 026402 (2018)}.

\bibitem{38} X. Dong, M. Hu, J. He, Y. Tian, and H.-T. Wang, A new phase from compression of carbon nanotubes with anisotropic Dirac fermions, {\color{blue}Sci. Rep. \textbf{5}, 10713 (2015)}.

\bibitem{39} C. Zhong, Y. Chen, Y. Xie, S. A. Yang, M. L. Cohen,and S. Zhang, Towards three-dimensional Weyl-surface semimetals in graphene networks, {\color{blue}Nanoscale \textbf{8}, 7232 (2016)}.

\bibitem{40} Y.-J. Jin, Z.-J. Chen, B.-W. Xia, Y.-J. Zhao, R. Wang, and H. Xu, Ideal intersecting nodal-ring phonons in bcc C8, {\color{blue}Phys. Rev. B. \textbf{98}, 220103(R) (2018)}.

\bibitem{41} Q.-B. Liu, Z.-Q. Wang, and H.-H. Fu, Topological phonons in allotropes of carbon, {\color{blue}Mater. Today Phys. \textbf{22}, 100694 (2022)}.

\bibitem{41_1} Supplementary materials in XXXX



\bibitem{42} R. Hoffmann, A. A. Kabanov, A. A. Golov, D. M. Proserpio, Homo Citans and Carbon Allotropes: For an Ethics of Citation, {\color{blue}Angew. Chem. Int. Ed. \textbf{55}, 10962–10976 (2016)}.

\bibitem{43} W. Kohn and L. J. Sham, Self-consistent equations including exchange and correlation effects, {\color{blue}Phys. Rev. B \textbf{136}, A1133 (1964)}.

\bibitem{44} P. E. Bl\"{o}chl, Projector augmentad-wave method, {\color{blue}Phys. Rev. B \textbf{50}, 17953 (1994)}.

\bibitem{45} J. P. Perdew, K. Burke, and M. Ernzerhof, Generalized gradient approximation made simple, {\color{blue}Phys. Rev. Lett. \textbf{77}, 3865 (1996)}.

\bibitem{46} A. Togo and I. Tanaka, First principles phonon calculations in materials science, {\color{blue}Scr. Mater. \textbf{108}, 1 (2015)}.

\bibitem{47} Q.-S. Wu, S.-N. Zhang, H.-F. Song, M. Troyer, and A. A. Soluyanov, WannierTools: an open-source software package for novel topological materials, {\color{blue}Comput. Phys. Commun. \textbf{243}, 110 (2019)}.

\bibitem{48} M. P. L. Sancho, J. M. L. Sancho, J. M. L. Sancho, and J. Rubio, Highly convergent schemes for the calculations of bulk and surface Green functions, {\color{blue}J. Phys. F: Metal Phys., \textbf{15}, 851 (1985)}.

\bibitem{49} R. Yu, X.-L. Qi, A. Bernevig, Z. Fang, and X. Dai, Equivalent expression of $Z_2$ topological invariant for band insulators using the non-abelian Berry connection. {\color{blue}Phys. Rev. B \textbf{84}, 075119 (2011)}.

\bibitem{50} J.-C. Gao, Q.-S. Wu, C. Persson, and Z. Wang, Irvsp: To obtain irreducible representations of electronic states in the VASP, {\color{blue}Comput. Phys. Commun., \textbf{261}, 107760 (2021)}.


\bibitem{51} Z.-M. Yu, Z. Zhang, G.-B. Liu, W. Wu, X.-P. Li, R.-W. Zhang, S.-A. Yang, and Y. Yao, Encyclopedia of emergent particles in three-dimensional crystals, {\color{blue}Sci. Bull. \textbf{10}, 023 (2022)}.

\bibitem{52} C. J. Bradley and A. P. Cracknell, The mathematical theory of symmetry in solids: representation theory for point groups and space groups, {\color{blue}Oxford University Press (2009)}.


\end{thebibliography}
\end{document}